\documentstyle[preprint,epsfig,aps]{revtex}

\begin{document}
\draft
\tighten
\title{Fully-Renormalized QRPA fulfills Ikeda sum rule exactly}
\author{Vadim Rodin and Amand Faessler}
\address{
Institut f\"ur Theoretische Physik der Universit\"at T\"ubingen,\\
Auf der Morgenstelle 14, D-72076 T\"ubingen, Germany\\}
\date{\today}
\maketitle
\begin{abstract}
The renormalized quasiparticle-RPA is reformulated for even-even nuclei
using restrictions imposed by the commutativity of the phonon
creation operator with the total particle number operator.
This new version,
Fully-Renormalized QRPA (FR-QRPA), is free from the spurious low-energy
solutions. Analytical proof is given that the Ikeda sum rule is
fulfilled within the FR-QRPA.
\end{abstract}
\pacs{21.60.Jz,  % HF and RPA
23.40.Hc % beta decay; double beta decay; Relation with nuclear matrix elements and nuclear structure
 }

The well-known Random Phase Approximation (RPA)~\cite{book} has found
many applications in
different fields of physics, from chemistry and condensed
matter~\cite{solstate} to relativistic field theory~\cite{PhTh}.
In nuclear physics RPA has been extensively exploited %used
to model properties of the excited nuclear states
that allows to calculate intensities of
various nuclear reactions, including
 %such an important one as
the double beta decay (see reviews \cite{fae98}).

The neutrinoless double beta decay ($0\nu\beta\beta$-decay),
which violates the total lepton number by two units,
is a sensitive low-energy probe for new physics beyond the
Standard Model~\cite{fae98,hax84}. The
observation of the $0\nu\beta\beta$-decay
would give unambiguous evidence that at least one of the
neutrinos is a Majorana particle with non-zero mass~\cite{sch82}.
The current experimental
upper limits on the $0\nu\beta\beta$-decay half-life impose stringent
constraints, e.g.,  on  the parameters
of Grand Unification and supersymmetric extensions
of the Standard Model.

It is important that the same nuclear structure methods can be applied to
calculate intensities of $2\nu\beta\beta$-decay, which is a second
order process allowed within the Standard Model. In this case the
quality of the RPA calculations can be directly checked
by comparison with the corresponding experimental data available at
the moment for a number of nuclei~\cite{exp}. Most of the nuclei are
open-shell ones and one uses the quasiparticle version of RPA (QRPA)
which takes into account nucleon pairing correlations.

The
 %performed analysis has shown that
QRPA is able to reproduce the experimental data~\cite{vog86}, but with
the strength of a particle-particle interaction which is
very close to the point where QRPA collapses. This well-known drawback of
the RPA is due to overestimation of ground state correlations
when one uses the quasiboson approximation (QBA) for the
bifermionic operators and the Pauli exclusion principle (PEP) is violated.
The proximity to the collapse makes the entire QRPA scheme unstable and
calculation results unreliable. To cure this problem, the so-called
renormalized QRPA (RQRPA) was
invented~\cite{rqrpa}, which takes into account PEP
in an approximate way. Namely, the expectation values of the
quasiparticle number operators are consistently treated to be
non-vanishing in the RQRPA vacuum. The RQRPA does not collapse
for physical values of the particle-particle interaction strength.
It has been shown within schematic models that by
including PEP in the QRPA good agreement with the exact solution of
the many-body problem can be achieved even beyond the critical point
of the standard QRPA (see, e.g. \cite{schm} and references therein).
The RQRPA has been used in previous studies
of the double beta decay \cite{fae98,FKSS97,Toi97}.

Nevertheless, it has been shown within the modern versions of RQRPA
that the model-independent Ikeda sum rule (ISR) is violated~\cite{Toi97,Sto01,Bob00}.
The Ikeda sum rule states that the difference between the total
Gamow-Teller strengths in the $\beta^-$ and $\beta^+$ channels is
$3(N-Z)$~\cite{Ikeda}.
This drawback of the model is quite serious, because the derivation of
the ISR is based just upon two principles: closure relation and
average conservation of the particle number. 
%: $\langle 0|\hat N |0\rangle=N$.
Violation of the ISR immediately means
violation of one (or both) of these important principles.
The drawback has not been cured neither in the self-consistent
QRPA~\cite{Del97} 
or in second-QRPA~\cite{Sto01}. The restoration of the ISR claimed  
in the Ref.~\cite{Del97} actually works only
for a special one-level model and can not be generalized for the
realistic case as shown in Ref.~\cite{Bob00}. At the moment, the
second-QRPA is able
to reduce the violation of the ISR to the level of few percents only
when contributions from the three (and more) boson states are neglected in the
boson expansion of the bifermionic operators~\cite{Sto01}.  

It can be conjectured that the closure relation is not fulfilled within the
standard formulation of the RQRPA. The vacuum of the RQRPA
contains quasiparticles.
The RQRPA takes into account the fact that the quasiparticles block
creation of the bifermionic QRPA bosons but
does not consider the possibility, that the quasiparticles in the
ground state take part in the transitions.
This leads to violation of the unitarity and underestimates
the ISR which should be fulfilled exactly.

In view of these remarks, modification of the phonon operator in order
to include so-called scattering terms (describing transitions of the
quasiparticles) is unavoidable if one wants to restore the ISR within
RQRPA.
This has been understood for some time and attempts have been
undertaken to treat the terms as additional independent constituents
of the phonon operator~\cite{rad98},
in spirit of the thermal extensions of RPA~\cite{tan88}
for description of the particle-hole excitations in heated nuclei.
That leads to increasing of the dimension of the QRPA equation system and
the appearance of low-lying energy roots originating from the transitions of the
quasiparticles.
 %with unperturbated energy $E_\pi-E_\nu$ (in exact
 % accordance with the above-mentioned analogy of the RQRPA and an odd
 % nucleus ground states).
Although such a consideration has allowed to restore the ISR for finite
temperatures~\cite{Ci00}, the low-lying spectrum has no physical
interpretation when the even-even nuclei are considered.
For instance,
the model would predict a number of $2^+$ excitations below the
first quadrupole state, that have never been observed in experiments
(see also the discussion in Ref.~\cite{Zelev01}).
Authors of a recent paper~\cite{Zelev01} have succeeded in removing
these spurious states by introducing new quasiparticles but they
have again used the QBA in terms of the new
quasiparticles and formulated the QRPA with respect to the HFB vacuum.

The above-mentioned examples show that a correct formulation
of the RQRPA is quite a delicate task.
 The goal of this Letter is to introduce a version of the
 RQRPA for even-even nuclei which is able to overcome mentioned
difficulties, i.e. it fulfills
the ISR and at the same time is free from the spurious solutions.

 Let us start from the general basis of the RPA, which assumes that
an excited state of the nucleus in question,
with the angular momentum $J$ and the projection $M$,
is created by applying the phonon-operator $Q^{\dagger}_{JM}$
to the vacuum state $|0^+_{RPA}\rangle$ of the initial, even-even, nucleus
with the same nucleon number $A$:
\begin{equation}
|JM  \rangle = Q^{\dagger}_{JM }|0^+_{RPA}\rangle
 \qquad \mbox{with} \qquad
Q_{JM }|0^+_{RPA}\rangle=0.
%\label{eq:19}
\end{equation}
At this point one usually introduces the following ansatz for
the phonon-operator $Q^{\dagger}_{JM }$:
\begin{equation}
Q^{\dagger}_{JM} = \sum_{\tau\tau'}
 \left [ X_{(\tau\tau', J )} A^\dagger(\tau\tau', JM)
- Y_{(\tau\tau', J )}\tilde{A}(\tau\tau', JM)\right ].
\label{Qa}
\end{equation}
where $X_{(\tau\tau',J)}$, $Y_{(\tau\tau',J)}$ denotes free
variational amplitudes, which are calculated by solving
the RQRPA equations;
$ A^\dagger(\tau\tau', JM)$  and $ A^{}(\tau\tau', JM)$
($\tau$ and $\tau'$ denote single-particle quantum numbers including
isospin projection) are the two quasi-particle creation and
annihilation operators:
\begin{eqnarray}
\label{Qaa}
A^\dagger(\tau\tau', JM) =
 %{\cal N}_{\tau \tau'}
\left[a^\dagger_{\tau} a^\dagger_{\tau'}\right]_{JM}
 %\\\left[a^\dagger_{\tau} a^\dagger_{\tau'}\right]_{JM}
&\equiv &\sum\limits^{}_{m_\tau , m_\tau'}
C^{J M}_{j_\tau m_\tau j_{\tau'} m_{\tau'} }
a^\dagger_{\tau m_\tau} a^\dagger_{\tau' m_\tau'}. 
\end{eqnarray}
The quasiparticle creation and annihilation operators
$a^{+}_{\tau m_\tau}$ and $a^{}_{\tau m_\tau}$
are defined by the Bogoliubov
 %-Valatin
transformation
\begin{equation}
\left( \matrix{ a^{+}_{\tau m_{\tau} } \cr
 {\tilde{a}}_{\tau  m_{\tau} }
}\right) = \left( \matrix{
u_{\tau} & v_{\tau} \cr
-v_{\tau} & u_{\tau}
}\right)
\left( \matrix{ c^{+}_{\tau m_{\tau}} \cr
{\tilde{c}}_{\tau m_{\tau}}
}\right),
\label{uv}
\end{equation}
where $c^{+}_{\tau m_\tau}$ ($c^{}_{\tau m_\tau}$)
denotes the particle creation (annihilation) operator.
%for a particle on a level with quantum numbers
%$(n_\tau, l_\tau, j_\tau )$.
The
 %parameters $u$, $v$
 %are occupation amplitudes and
 %the
tilde symbol indicates the  time-reversal operation, e.g.
$\tilde{a}_{\tau {m}_{\tau}}$ =
$(-1)^{j_{\tau} - m_{\tau}}a^{}_{\tau -m_{\tau}}$.

However it turns out, that the most appropriate way is to
analyze the phonon structure in terms of the particle
creation and annihilation operators (instead of the quasiparticle ones).
Let us
first mention two quite general conditions which should be kept in mind
while constructing the phonon operator. Namely,
because the phonon operator is assumed to create a nuclear state of $A$
nucleons acting on the $0^+$ ground state of the initial nucleus (also of $A$
nucleons), the operator has to
\begin{itemize}
\item[ 1)] have good angular momentum $J$ and the projection $M$ (it has been
 taken into account in Eqs.~(\ref{Qa}),(\ref{Qaa}));\\
\item[ 2)] commute with the total particle number
    operator $\hat A =\hat N + \hat Z=
\sum\limits^{}_{\tau , m_\tau}c^\dagger_{\tau m_\tau}{{c}}_{\tau m_\tau}
  %=-\sum\limits^{}_{\tau}\hat j_\tau[c^\dagger_{\tau}{\tilde{c}}_{\tau}]_{00}, \ \ (\hat j=\sqrt{2j+1})
$.
Neutral excitations can be further distinguished from the
charge-exchange ones 
using their commutativity with $\hat N - \hat Z$.
\end{itemize}

The first rule tells us that the simplest building blocks for the phonon operator
are $\left [c^\dagger_{\tau}{\tilde{c}}_{\tau'}\right]_{JM}$, $\left[c^\dagger_{\tau}c^\dagger_{\tau'}\right]_{JM}$ together
with their hermitian conjugates.
 %and $[c_{1}c_{2}]_{JM}$ (without specifying yet the isospin indices).
Applying the second rule leads to a further restriction of the phonon
operator structure:
 %for a charge-exchange phonon:
\begin{equation}
Q^{\dagger}_{JM } = \sum_{\tau\tau'}
 \left [x_{(\tau\tau', J )} C^\dagger(\tau\tau', JM)
-y_{(\tau\tau', J )}\tilde{C}(\tau\tau', JM)\right ].
\label{Qc}
\end{equation}
with $C^\dagger(\tau\tau', JM)=\left[c^\dagger_{\tau}{\tilde{c}}_{\tau'}\right]_{JM}$ and
 $\tilde C(\tau\tau', JM)=(-)^{J-M}C(\tau\tau', J\,-M)$. For the neutral excitations
$\tau$ and $\tau'$ have the same isospin projections,
whereas for the charge-exchange ones - opposite.
Therefore, in terms of the original particle creation and annihilation
operators $Q$ consists of scattering terms only.
Going now into the quasiparticle representation, one gets:
\begin{eqnarray}
C^\dagger(\tau\tau', JM) &=&
u_{\tau} v_{\tau'} A^\dagger(\tau\tau', JM) + u_{\tau'} v_{\tau} \tilde{A} (\tau\tau', JM)
\nonumber\\
&+&
u_{\tau} u_{\tau'} B^\dagger(\tau\tau', JM) - v_{\tau} v_{\tau'} \tilde{B} (\tau\tau', JM)
\nonumber\\
&=&
u_{\tau} v_{\tau'} \bar A^\dagger(\tau\tau', JM) + v_{\tau} u_{\tau'} \tilde{\bar A\,}(\tau\tau', JM).
\label{Cdagger}
\end{eqnarray}
with the following definitions:
$B^\dagger(\tau\tau', JM)=\left[a^\dagger_{\tau}{\tilde{a}}_{\tau'}\right]_{JM}$
for the quasiparticle scattering term and
$\bar A^\dagger= A^\dagger+\left(u_{\tau'}v_{\tau'}B^\dagger-
u_{\tau}v_{\tau}\tilde B\right)\left/(v_{\tau'}^2-v_{\tau}^2)\right.$ for
the bifermionic operator $\bar A^\dagger$.

Now we are able to rewrite (\ref{Qc}) in the form which is similar to (\ref{Qa}):
\begin{equation}
Q^{\dagger}_{JM } = \sum_{{\tau}{\tau'}}
 \left [ X_{({\tau}{\tau'}, J )} \bar A^\dagger({\tau}{\tau'}, JM)
- Y_{({\tau}{\tau'}, J )}\tilde{\bar A\,}({\tau}{\tau'}, JM)\right ],
\label{Qa1}
\end{equation}
where $X=u_{\tau}v_{\tau'}x-v_{\tau}u_{\tau'}y,
\ Y=u_{\tau}v_{\tau'}y-v_{\tau}u_{\tau'}x$.
Thus, instead of $A^\dagger, A$, bifermionic operators $\bar
A^\dagger,\bar A$ are now the basic terms in which
the RQRPA should be formulated. The key
point is that the latter automatically
contain the quasiparticle scattering terms which, however, are not associated
with any additional degrees of freedom. That means that there are no
spurious low-lying solutions in the present theoretical scheme.

From this point we can follow the usual way to formulate the RQRPA~\cite{rqrpa}
substituting everywhere $A$ by $\bar A$.
The forward- and backward- going free variational amplitudes X and Y
satisfy the equation:
\begin{equation}
\left(
\begin{array}{cc}
{\cal A}&{\cal B}\\
{\cal B}&{\cal A}
\end{array}
\right)
\left(
\begin{array}{c}
X\\
Y
\end{array}
\right)
= {\cal E}_{QRPA}
\left(
\begin{array}{cc}
{\cal U}&0\\
0&{\cal -U}
\end{array}
\right)
\left(
\begin{array}{c}
X\\
Y
\end{array}
\right),
 %\label{dqrpa)
%\label{eq:13}
\end{equation}
where
\begin{eqnarray}
{\cal A} &=& \langle 0^+_{RPA}| \left[ \bar A, \left[ H_F, \bar
A^\dagger \right ]\right]
|0^+_{RPA} \rangle, \nonumber \\
{\cal B} &=& - \langle 0^+_{RPA}| \left[ \bar A, \left[ H_F, \bar
A \right ]\right] |0^+_{RPA}
\rangle,
 %{\cal U} &=&  <rpa| [\bar A,\bar A^\dagger ] |rpa>.
\label{eq:14}
\end{eqnarray}
$H_F$ is the nuclear Hamiltonian and the renormalization matrix
 %${\cal U}_{p n}$ and
${\cal U}_{\tau\tau'}$ is
%($\tau=p,n$) are:
%the following expectation values:

\begin{eqnarray}
 %{\cal U}_{p n}=\langle 0^+_{RPA}|[\bar A(p n, JM),
 %\bar A^\dagger(p'n', JM)]|0^+_{RPA}\rangle &=&
 %\delta_{p p'}\delta_{n n'}
 %{\cal D}_{p n},
   %\nonumber \\
{\cal U}_{\tau\tau'}&=&\langle 0^+_{RPA}|\left[ \bar A(\tau\tau', JM),
\bar A^\dagger(\sigma\sigma', JM)\right]|0^+_{RPA}\rangle =
   %(
\delta_{\tau\sigma}\delta_{\tau'\sigma'}
   % -(-1)^{j_\tau + j_{\tau'} -J} \delta_{\tau\sigma'}\delta_{\tau'\sigma})
{\cal D}_{\tau\tau'},\nonumber\\
{\cal D}_{\tau\tau'}&=&1+\left((u_{\tau'}^2-v_{\tau'}^2){\cal  N}_{\tau'}
-(u_{\tau}^2-v_{\tau}^2){\cal N}_{\tau}\right)\left/
(v_{\tau'}^2-v_{\tau}^2)\right.
 %\left((u_{\tau}^2-v_{\tau}^2)\hat j_\tau^{-1}\langle
 %B(\tau\tau,00)\rangle\right. &-&\left.
 %(u_{\tau'}^2-v_{\tau'}^2)\hat j_{\tau'}^{-1}\langle
 % B(\tau'\tau',00)\rangle  \right)/(v_{\tau'}^2-v_{\tau}^2)
\label{Dm}
\end{eqnarray}
Here, the exact (fermionic) expressions of the commutators are taken
into account and
\begin{equation}
{\cal N}_{\tau}=\hat j_\tau^{-2}\langle0^+_{RPA}|\sum\limits^{}_{m_\tau}
 a^\dagger_{\tau m_\tau}{{a}}_{\tau m_\tau}|0^+_{RPA}\rangle
=-\hat j_\tau^{-1}\langle 0^+_{RPA}|B(\tau\tau,00)|0^+_{RPA}\rangle
\end{equation}
gives the relative
quasiparticle occupation number for the level $\tau$ in the RQRPA
vacuum ($\hat j\equiv\sqrt{2j+1}$). The method to calculate these
occupation numbers
%is out of scope of this paper and
can be found elsewhere~\cite{rqrpa}. It is useful to introduce the notation:
\begin{equation}
\bar{X} = {\cal U}^{1/2} X, ~~~~~~\bar{Y} = {\cal U}^{1/2} Y,
\label{eq:15}
\end{equation}
\begin{equation}
\bar{\cal A} = {\cal U}^{-1/2} {\cal A} {\cal U}^{-1/2}, ~~~~
\bar{\cal B} = {\cal U}^{-1/2} {\cal B} {\cal U}^{-1/2}.
\label{eq:16}
\end{equation}
Then the amplitudes $\bar X$ and $\bar Y$ satisfy the equation of usual QRPA:
\begin{equation}
\left(
\begin{array}{cc}
\bar {\cal A}&\bar {\cal B}\\
\bar {\cal B}&\bar {\cal A}
\end{array}
\right)
\left(
\begin{array}{c}
\bar X\\
\bar Y
\end{array}
\right)
= {\cal E}_{QRPA}
\left(
\begin{array}{cc}
1&0\\
0& -1
\end{array}
\right)
\left(
\begin{array}{c}
\bar X\\
\bar Y
\end{array}
\right),
%\label{dqrpa)
%\label{eq:13}
\end{equation}
%%%%%%%%%%%%%%%%%%%%%%%%%%%%%%%%%%%%%%%%%%%%%%
Solving the FR-QRPA  equations, one gets the fully
renormalized amplitudes $\bar{X}$, $\bar{Y}$
with the usual normalization and closure relation, which will be used
below to prove the fulfillment of the ISR:
\begin{eqnarray}
\sum\limits_{\tau\tau'}\bar{X}^m_{(\tau\tau', J )}\bar{X}^k_{(\tau\tau', J )}
-\bar{Y}^m_{(\tau\tau', J)}\bar{Y}^k_{(\tau\tau', J )}&=&\delta_{km},
\nonumber\\
\sum\limits_{m}\bar{X}^m_{(\tau\tau', J )}\bar{X}^m_{(\tau^{\phantom 1}_1\tau'_2, J )}
-\bar{Y}^m_{(\tau\tau', J )}\bar{Y}^m_{(\tau^{\phantom 1}_1\tau'_2, J )}&=&
\delta_{\tau\tau^{\phantom 1}_1}\delta_{\tau'\tau'_2},\nonumber\\
\sum\limits_{m}\bar{X}^m_{(\tau\tau', J )}\bar{Y}^m_{(\tau^{\phantom 1}_1\tau'_2, J )}
-\bar{Y}^m_{(\tau\tau', J )}\bar{X}^m_{(\tau^{\phantom 1}_1\tau'_2, J )}&=&0.
\label{closure}
\end{eqnarray}
Here, $m$ and $k$ mark different roots of the QRPA equations for a
given $J^\pi$.

Now we are ready to prove the fulfillment of the ISR within the
FR-QRPA. Let restrict ourselves to the charge-exchange
pn-FR-QRPA and write down the definition for the ISR:
\begin{equation}
ISR=\sum\limits_m \left|\langle JM,  m| \beta_{JM}^\dagger | 0^+_{RPA}\rangle
\right|^2- \sum\limits_{m'}
\left|\langle J\, -M, m'| \beta_{JM} | 0^+_{RPA}\rangle \right|^2.
\label{ISR}
\end{equation}
Here, we would like to consider the Fermi and Gamow-Teller transitions
 simultaneously by defining:
\begin{equation}
\beta_{JM}^\dagger=\sum\limits_{pn, m_pm_n}
\langle pm_p| q_{JM}| nm_n\rangle c^\dagger_{p}{c}_{n}=
-\hat J^{-1}
\sum\limits_{pn}\langle p\| q_{J}\| n\rangle C^\dagger(pn, JM),
\end{equation}
where $q_{00}=1$ (Fermi transitions) and $q_{1\mu}=\sigma_{\mu}$
(Gamow-Teller transitions) and one has for both cases $ISR=N-Z$ (in
the latter case -  just because of the only fixed
angular momentum projection).
One can use Eqs.~(\ref{Cdagger}),(\ref{Qa1}) to get:
\begin{eqnarray}
\langle JM,  m| \beta_{JM}^\dagger | 0^+_{RPA}\rangle&=&-\hat J^{-1}
\sum_{pn} \langle p\|q_{J} \| n\rangle
\left(u_{p} v_{n} {\bar{X}}^{m}_{(pn, J )}
+v_{p} u_{n} {\bar{Y}}^{m}_{(pn, J )}\right)
\sqrt{{\cal D}_{pn}},
\nonumber\\
\langle J\,-M,  m| \beta_{JM}| 0^+_{RPA}\rangle&=&-(-)^{J-M}\hat J^{-1}
\sum_{pn} \langle n\|q_{J} \| p\rangle
\left (u_{p} v_{n} {\bar{Y}}^{m}_{(pn, J )}
+v_{p} u_{n} {\bar{X}}^{m}_{(pn, J )}\right)
\sqrt{{\cal D}_{pn}}.
\label{me}
\end{eqnarray}
Substituting (\ref{me}) into (\ref{ISR}) and making
use of the closure conditions (\ref{closure}), one ends up with
\begin{equation}
ISR=\hat J^{-2}
\sum_{pn} \left|\langle p\|q_{J} \| n\rangle\right|^2
(v_n^2-v_p^2) {\cal D}_{pn}.
\end{equation}

Within the usual QBA one has ${\cal D}_{pn}=1$ and 
\begin{equation}
ISR= \hat J^{-2}\left(
\sum\limits_{np} \left|\langle p\|q_{J} \| n\rangle\right|^2
v_n^2 -\sum\limits_{np}
\left|\langle p\|q_{J} \| n\rangle\right|^2 v_p^2\right)=N-Z, 
\end{equation}
making use of \ $\sum\limits_{p}\left|\langle p\|q_{J}
\|n\rangle\right|^2 = \hat J^2 \hat j_n^2,\
\sum\limits_{n}\left|\langle p\|q_{J}
\|n\rangle\right|^2 = \hat J^2 \hat j_p^2$ 
and of the equations for the BCS chemical potentials $\lambda_n,\lambda_p$ 
\begin{eqnarray}
\langle 0^+_{HFB}|\hat N
|0^+_{HFB}\rangle=\sum\limits_{n}\hat j_n^2 v_n^2=N, \ \ \
\langle 0^+_{HFB}|\hat Z
|0^+_{HFB}\rangle=\sum\limits_{p}\hat j_p^2 v_p^2=Z.
\end{eqnarray}
 %along with the same equations for the proton subsystem.
This corresponds to the fulfillment of the ISR for the usual QRPA.

In the
case of the FR-QRPA with ${\cal D}_{pn}$ determined by Eq.~(\ref{Dm})
one has
\begin{eqnarray}
ISR&=&
\sum\limits_{n} \hat j_n^2\left(v_n^2+(u_n^2-v_n^2){\cal N}_{n}
 %-\hat j_n^{-1}(u_n^2-v_n^2)\left<B(nn,00) \right>
\right)\nonumber\\
&-&\sum\limits_{p}
\hat j_p^{2}\left(v_p^2+(u_p^2-v_p^2){\cal N}_{p}
 %-\hat j_p^{-1}(u_p^2-v_p^2)\left<B(pp,00)\right>
\right)\nonumber\\
&=&N-Z
\end{eqnarray}
because of the modified FR-QRPA equations for the chemical potentials:
\begin{eqnarray}
\langle 0^+_{RPA}|\hat N |0^+_{RPA}\rangle&=&\sum\limits_{n}
\hat j_n^{2}\left(v_n^2+(u_n^2-v_n^2){\cal N}_{n}\right)=N
 %-\hat j_n^{-1}(u_n^2-v_n^2)\left<B(nn,00)\right>
,\nonumber\\
\langle 0^+_{RPA}|\hat Z|0^+_{RPA}\rangle&=&\sum\limits_{p}
\hat j_p^{2}\left(v_p^2+(u_p^2-v_p^2){\cal N}_{p}\right)=Z.
\end{eqnarray}

In conclusions, we have reformulated the RQRPA into Fully-Renormalized
QRPA (FR-QRPA) for even-even nuclei
using restrictions imposed by the commutativity of the phonon
creation operator with the total particle number operator.
The FR-QRPA is free from the spurious low-energy solutions.
We have also shown analytically that the Ikeda sum rule is fulfilled
within the FR-QRPA.

\acknowledgements
V.R. would like to thank the Graduiertenkolleg "Hadronen im Vakuum,
in Kernen und Sternen" GRK683 for supporting his stay in T\"ubingen.


\begin{references}
\bibitem{book}
A.L. Fetter and J.D. Walecka, {\it Quantum Theory of Many-Particle
Systems} (McGraw-Hill, New York, 1971);
A. Bohr and B. Mottelson, {\it Nuclear structure} (Benjamin, New York,
1975) Vol.II; P. Ring and P. Schuck, {\it The Nuclear Many Body Problem}
(Springer-Verlag, Berlin, 1980).
\bibitem{solstate}
P. Krueger, P. Schuck, Europhys. Lett. 72, 395 (1994);
F. Catara, G. Piccitto, M. Sambataro, N. Van Giai
  Phys. Rev. {\bf B 54}, 17536 (1996);
S. Schafer, P. Schuck, Phys. Rev. {\bf B 59}, 1712 (1999).
\bibitem{PhTh} T. Bertrand, P. Schuck, G. Chanfray, Z. Aouissat, J. Dukelsky
Phys. Rev. {\bf C 63}, 024301 (2001),
 H. Hansen, G. Chanfray, D. Davesne, P. Schuck, hep-ph/0201279
%
%%%%
% dbd reviews
%%%%
\bibitem{fae98} A. Faessler and F. \v Simkovic,
  J. Phys. G {\bf  24}, 2139 (1998);
J. Suhonen and O. Civitarese, Phys. Rep. {\bf 300}, 123 (1998).

\bibitem{hax84} W.C. Haxton and G.J. Stephenson, Progr. Part.
                Nucl. Phys. {\bf 12}, 409 (1984);
                J. D. Vergados, Phys. Report, {\bf 133}, 1 (1986);
        K. Grotz and H.V. Klapdor-Kleingrothaus, {\it The Weak Interactions
        in Nuclear, Particle and Astrophysics}
        (Adam Hilger, Bristol, New York, 1990);
        R.N. Mohapatra and P.B. Pal, {\it Massive Neutrinos in Physics
                 and Astrophysics} (World Scientific, Singapore, 1991);
 %\bibitem{doi85} 
M. Doi, T. Kotani and E. Takasugi, Progr. Theor. Phys.
                        Suppl. {\bf 83}, 1  (1985);
 %\bibitem{pv} 
M. Moe, P. Vogel, Ann. Rev. Nucl. Part. Sci. {\bf 44},
247 (1994); P. Vogel, nucl-th/0005020; S. R. Elliot, P. Vogel
hep-ph/0202264.
%
\bibitem{sch82} J. Schechter and J.W.F. Valle, Phys. Rev. {\bf D {25}}, 
{774} (1982);
 M. Hirsch, H.V. Klapdor-Kleingrothaus, S.G. Kovalenko, 
Phys. Lett. B 398, 311 (1997).
\bibitem{exp} V.I. Tretyak, Yu.G.Zdesenko, At.Data Nucl.Data Tables
{\bf 80}, 83 (2002).
%%%%%%%%%%%%%%%
\bibitem{vog86} P. Vogel and M.R. Zirnbauer, Phys. Rev. Lett. {\bf 57}, 3148 (1986);
J. Engel, P. Vogel, and M. R. Zirnbauer, Phys. Rev. C 
 {\bf 37}, 731 (1988);
O. Civitarese, A. Faessler, and T. Tomoda, Phys. Lett.
   B {\bf 194}, 11 (1987);
K. Muto, E. Bender, and H. V. Klapdor, Z. Phys. A 
   {\bf 334}, 177 (1989).
\bibitem{rqrpa} K. Hara, Prog. Theor. Phys. {\bf 32}, 88 (1964);
D.J. Rowe, Rev. Mod. Phys. {\bf 40}, 153 (1968);
D. Karadjov, V.V. Voronov, and F. Catara, Phys. Lett. B
   {\bf 306}, 197 (1993);
F. Catara, N. Dinh Dang and M. Sambataro, Nucl. Phys. A
  {\bf 579}, 1 (1994).
%%%%%%%%%%%%%%%%%%
\bibitem{schm} F. \v Simkovic, A.A. Raduta, M. Veselsk\'y, and A. Faessler,
   Phys. Rev. {\bf C 61 }, 044319 (2000).
%%%%%%%%%%%%%%%%%
\bibitem{FKSS97} J. Toivanen and J. Suhonen, Phys. Rev. Lett. 
  {\bf 75}, 410 (1995); A. Faessler, S. Kovalenko, F. \v Simkovic, 
   and J. Schwieger, Phys. Rev. Lett. {\bf 78}, 183  (1997); 
   A. Faessler, S. Kovalenko, and F. \v Simkovic,
   Phys. Rev. {\bf D 58}, 115004  (1998);
J. Schwieger, F. \v Simkovic, A. Faessler, W.A. Kami\'nski.
   Phys. Rev. {\bf C 57}, 1738 (1998);
F. \v Simkovic, G. Pantis, J.D. Vergados, and A. Faessler,
  Phys. Rev. {\bf C 60}, 055502 (1999);
A.A. Raduta, F. \v Simkovic, A. Faessler,
    J. Phys. {\bf G 26}, 793 (2000);
F. \v Simkovic, N. Nowak, W.A. Kami\'nski,
A.A. Raduta and A. Faessler, Phys. Rev. {\bf C 64}, 035501 (2001).
 % violation of ISR in RQRPA
\bibitem{Toi97} J. Toivanen and J. Suhonen, Phys. Rev. {\bf C 55},
2314 (1997).
\bibitem{Sto01} S. Stoica and H.V. Klapdor-Kleingrothaus,
 Eur. Phys. J. A {\bf 9}, 345 (2000); Phys. Rev. {\bf C 63}, 064304
(2001); Nucl. Phys. A {\bf 694}, 269 (2001).
 %S. Stoica, I. Mihut, and J. Suhonen, Phys. Rev. {\bf C 64},
 %017303 (2001).
\bibitem{Bob00} A. Bobyk, W.A. Kaminski, P. Zareba, Nucl. Phys. A
{\bf 669}, 221 (2000); P. Zareba, PhD thesis, 2000, unpublished.
\bibitem{Ikeda} K. Ikeda, Prog. Theor. Phys. {\bf 31}, 434 (1964).
 %K. Ikeda, T. Udagawa, H. Yamamura, Prog. Theor. Phys. 33 (1965) 22
\bibitem{Del97} D.S. Delion, J. Dukelsky and P. Schuck,
   Phys. Rev. {\bf C 55 }, 2340 (1997).
 % RQRPA with B,B^\dagger
\bibitem{rad98} A.A. Raduta, C.M. Raduta, A. Faessler, and W. Kami\'nski,
  Nucl. Phys. A {\bf 634}, 497 (1998); 
N.~D.~Dang and A.~Arima, Phys. Rev. {\bf C 62 }, 024303 (2000).
\bibitem{tan88} K. Tanabe, Phys. Rev. {\bf C 37}, 2802 (1988);
 T.~Hatsuda, Nucl. Phys. A {\bf 492}, 187 (1989).
\bibitem{Ci00} O. Civitarese, J.G. Hirsch, F. Montani and M. Reboiro,
   Phys. Rev. {\bf C 62 }, 054318 (2000);
   O.~Civitarese and M.~Reboiro, Phys. Rev. {\bf C 63 }, 034323 (2001).
\bibitem{Zelev01} N. Dinh Dang and V. Zelevinsky, Phys. Rev. {\bf C 64
}, 064319 (2001).
%
\end{references}
\end{document}